\documentclass[11pt,aps,prd,superscriptaddress,amsmath,amssymb,nofootinbib]{revtex4-2}
\usepackage[T1]{fontenc}
\usepackage[utf8]{inputenc}
\usepackage[colorlinks=true, allcolors=blue]{hyperref}
\usepackage{graphicx}
\newcounter{errata}
\alph{errata}
\begin{document}


\title{Primordial perturbations and inflation in holographic cosmology}
\author{Nicolas R.\ Bertini}\email{nicolas.bertini@cosmo-ufes.org}
\affiliation{Núcleo de Astrofísica e Cosmologia \& Departamento de F\'isica, Universidade 
Federal do Esp\'irito Santo. Av.\ Fernando Ferrari 514, 29075-910, Vit\'oria, ES, Brazil}
\author{Neven Bili\'c}\email{bilic@irb.hr}
\affiliation{Núcleo de Astrofísica e Cosmologia \& Departamento de F\'isica, Universidade 
Federal do Esp\'irito Santo. Av.\ Fernando Ferrari 514, 29075-910, Vit\'oria, ES, Brazil}
\affiliation{Division of Theoretical Physics, Rudjer Bo\v skovi\'c Institute, 10002 Zagreb, Croatia}
\author{Davi C.\  Rodrigues}\email{davi.rodrigues@cosmo-ufes.org}
\affiliation{Núcleo de Astrofísica e Cosmologia \& Departamento de F\'isica, Universidade 
Federal do Esp\'irito Santo. Av.\ Fernando Ferrari 514, 29075-910, Vit\'oria, ES, Brazil}

\begin{abstract}
We consider an inflationary scenario in the holographic braneworld with
a cosmological fluid occupying the 3+1 dimensional brane located at the holographic boundary
of an asymptotic ADS$_5$ bulk.
The contribution of the boundary conformal field can be represented as
a modification of Einstein's equations on the boundary.
Using these effective Einstein equations we calculate the cosmological perturbations
and derive the corresponding power spectra assuming a general $k$-essence type of
inflaton. 
We find that the braneworld scenario affects the scalar power spectrum only in the speed of sound
dependence on the slow-roll parameters whereas there is no change in the 
tensor power spectrum. This implies that the changes in 
the spectral indices appear at the second order  in the slow-roll parameter expansion.
\end{abstract}

\maketitle

\section{Introduction}

The anti-de Sitter/conformal field theory (AdS/CFT) correspondence  \cite{maldacena,gubser,witten1} 
goes beyond pure string theory and  links
many important theoretical and phenomenological issues.
 In particular, a simple physically relevant model related to  AdS/CFT
is the Randall-Sundrum (RS) braneworld model \cite{randall1,randall2} and its cosmological applications.
In the braneworld scheme, the RS brane provides a cutoff regularization
for the infrared divergences of the on shell bulk action \cite{gubser2,nojiri1, giddings,deharo,hawking,duff3,nojiri2,deharo2}.
A related scheme is based on the holographic braneworld scenario \cite{apostolopoulos,bilic1,nojiri,bilic3}
in which the cosmology is derived from the effective four-dimensional Einstein equations
on the holographic boundary of AdS$_5$.

In a braneworld  scenario, matter is confined
on a brane moving in the higher dimensional bulk,
with only gravity allowed to propagate in the bulk  \cite{arkani,antoniadis,maartens}.
If the brane is located at the boundary of
a 5-dimensional asymptotically AdS space-time,
we refer to this type of braneworld as the holographic braneworld
\cite{apostolopoulos,bilic1}. In the study of the holographic braneworld,  
a crucial property of an asymptotically AdS bulk is that  AdS space is dual to a conformal field theory at
its boundary.
A connection between AdS/CFT correspondence and cosmology has been studied in a different approach
based on a holographic renormalization group flows
in quantum field theory \cite{kiritsis2,binetruy}.

The holographic cosmology has a property that
the universe evolution starts from a point at which the energy density and cosmological scale
are both finite \cite{gao,bilic2},
rather than from the usual Big Bang singularity of the standard cosmology.
Then the inflation phase proceeds naturally immediately after $t=0$.
This type of inflation has been recently studied in
 a holographic braneworld scenario
with an effective tachyon field on the brane \cite{bilic3}.
In that paper the first order cosmological perturbations have been calculated in an approximate scheme 
in which the modification of Einstein equations were treated as an approximate modification of
the effective energy momentum tensor at the boundary. 
Here we study the complete first-order perturbation theory on the holographic brane
with field equations on the brane modified due to the dual conformal theory
on the AdS boundary. 

Our main motivation  is to show 
how the modification of Einstein's equation
on the holographic braneworld affects the primordial power spectra.
As we shall shortly demonstrate, the modification of Einstein equations 
on the holographic brane is substantial and yields a modified background cosmology
with a quartic term $\propto H^4$ in Friedmann equations, 
as shown previously in a different way \cite{apostolopoulos,bilic1}.
In view of that, one would expect  the first order perturbation to be
substantially modified yielding a modified power spectra. 
We assume that inflation is driven by a general $k$-essence \cite{armendariz1,armendariz2}.
The term  $k$-essence denotes a fluid described by a field theory with Lagrangian
that is a general function of the field and its kinetic term.
This form of field theory, which includes a canonical scalar field as a special case,
was first explored as a generalized scalar field model of inflation
\cite{armendariz,garriga}, with speed of sound being a nontrivial function of the field.
Study of a general $k$-essence is important as it covers many models of inflation (including canonical
scalar field inflation) with predictions in agreement with observations
(for recent works see \cite{saitou,barbosa,kamenshchik,lin,lola}).

The remainder of the paper is organized as follows.
In Sec.\ \ref{renormalization} we introduce the effective Einstein equations
on the holographic boundary from which we derive the corresponding Friedmann equation.
In Sec.\ \ref{perturbations} we study the primordial scalar and tensor perturbations
and calculate the power spectra and spectral indices. 
In Appendix \ref{appendix} we demonstrate equivalence of the energy-momentum conservation and the 
equation of motion for a general $k$-essence field theory. In Appendix \ref{appendixB} we directly verify 
the $k$-essence perturbations without using the fluid notation.
In Appendix \ref{appendixC} we briefly discuss  the background-field solutions for a few examples of $k$-essence.

\section{Holographic Einstein equations and cosmology}
\label{renormalization}

A general asymptotically AdS$_5$  metric in Fefferman-Graham coordinates
is of the form \cite{fefferman}
\begin{equation}
ds^2=G_{ab}dx^adx^b =\frac{\ell^2}{z^2}\left( g_{\mu\nu} dx^\mu dx^\nu -dz^2\right),
 \label{eq3001}
\end{equation}
where  the constant $\ell$ of dimension of length is
the AdS curvature radius.
We use the Latin alphabet for 4+1  and the Greek alphabet for 3+1 spacetime indices and
the metric sign convention is $(+----)$. 
Near $z=0$ the metric can be expanded as
\begin{equation}
  g_{\mu\nu}(z,x)=g^{(0)}_{\mu\nu}(x)+z^2 g^{(2)}_{\mu\nu}(x)+z^4 g^{(4)}_{\mu\nu}(x) +\cdots .
  \label{eq107}
 \end{equation}
Then, the four-dimensional Einstein equations on the holographic boundary are \cite{deharo,bilic1,bilic2} 
\begin{equation}
R_{\mu\nu}- \frac12 R g^{(0)}_{\mu\nu}= 8\pi G_{\rm N} (\langle T^{\rm CFT}_{\mu\nu}\rangle +T_{\mu\nu}),
 \label{eq3002}
\end{equation}
where $R$ and $R_{\mu\nu}$ are respectively the Ricci scalar and the Ricci tensor associated with the metric $g^{(0)}_{\mu\nu}$. 
The Newton's constant is related to the AdS curvature radius $\ell$ and  
five-dimensional gravitational constant $G_5$ as
\begin{equation}
 G_{\rm N} = \frac{2 G_5}{\ell} .
 \label{eq0001}
\end{equation}
In the following  analysis we shall consider $G_{\rm N}$  and $\ell$ as fixed fundamental  
physical parameters and $G_5$ as a derived quantity.

In general, on the left-hand 
side of (\ref{eq3002}) there is a cosmological term $\Lambda g_{\mu \nu}^{(0)}$ 
related to the brane tension and the AdS bulk cosmological constant
\cite{apostolopoulos,bilic1}.
This term 
can be  made small or entirely removed by  imposing the RS 
fine tuning condition \cite{bilic1} and 
if it were present would be important for the late Universe cosmology and 
irrelevant for our early universe considerations.

The AdS curvature radius $\ell$ is  constrained by tests of Newton's law at small distances.
For the Randall-Sundrum braneworld 
it has been shown
\cite{garriga2} that for $r\gg \ell$ 
the extra-dimension  effects 
strengthen Newton's gravitational field by a factor $1+\mathcal{O}(\ell^2/r^2)$. 
Table-top tests of Newton's laws currently find no deviations of Newton's potential
at distances greater than about 0.1 mm \cite{long} and about 0.05 mm \cite{lee} yielding an upper bound on the AdS$_5$
curvature
$\ell \lesssim 0.1$ mm or $\ell^{-1} \gtrsim 10^{-12}$ GeV.

The energy-momentum tensor $T_{\mu\nu}$
describes matter on the brane.
The vacuum expectation value $\langle T^{\rm CFT}_{\mu\nu}\rangle$  can be obtained in terms
of the quantities $g^{(2n)}_{\mu\nu}$ related to the bulk metric \cite{deharo}
\begin{eqnarray}
  \langle T^{\rm CFT}_{\mu\nu}\rangle=
- \frac{\ell^2}{2\pi G_{\rm N}}\left\{g^{(4)}_{\mu\nu}
-\frac18 \left[({\rm Tr} g^{(2)})^2-{\rm Tr} (g^{(2)})^2\right]g^{(0)}_{\mu\nu}
 -\frac12 (g^{(2)})^2_{\mu\nu}+\frac14 {\rm Tr} g^{(2)}g^{(2)}_{\mu\nu}
  \right\}.
 \label{eq3106}
\end{eqnarray}
Here
\begin{equation}
g^{(2)}_{\mu\nu}=\frac12 \left( R_{\mu\nu}-\frac16 R g^{(0)}_{\mu\nu}\right)
\end{equation}
 and
\begin{equation}
	g^{(4)}_{\mu\nu}=g^{\rm cf(4)}_{\mu\nu}+\tilde{g}^{(4)}_{\mu\nu},
\label{a1}
\end{equation}
where
\begin{equation}\tag{7a}
g^{\rm cf(4)}_{\mu\nu}=\frac14 (g^{(2)})^2_{\mu\nu} \equiv
\frac14 g^{(2)}_{\mu\rho}g^{(0)\rho\sigma}g^{(2)}_{\sigma\nu} .
\label{a}
\end{equation}
The tensor $\tilde{g}^{(4)}$ depends on the boundary metric $g^{(0)}_{\mu\nu}$ and vanishes if the metric is conformally flat.
Then, we can write 
\begin{equation}
  \langle T^{\rm CFT}_{\mu\nu}\rangle=
 -\frac{\ell^2}{32\pi G_{\rm N}}\left[\frac23 R R_{\mu\nu}-R_{\mu\rho}{R^\rho}_\nu
 +\frac14\left( 2R_{\alpha\beta}R^{\alpha\beta}
 - R^2\right)g^{(0)}_{\mu\nu} \right] +t_{\mu\nu}
  \label{eq4106}
\end{equation}
where
\begin{equation}\tag{8a}
t_{\mu\nu}=	-\frac{\ell^2}{2\pi G_{\rm N}}\tilde{g}^{(4)}_{\mu\nu} .
\end{equation}

From (\ref{eq3002}) and (\ref{eq4106}) we obtain the holographic Einstein equations
\begin{equation}
  R_{\mu\nu}- \frac12 R g^{(0)}_{\mu\nu}
 +\frac{\ell^2}{4}\left[\frac23 R R_{\mu\nu}-R_{\mu\rho}{R^\rho}_\nu
 +\frac14\left( 2R_{\alpha\beta}R^{\alpha\beta}
 - R^2\right)g^{(0)}_{\mu\nu} \right] 
 = 8\pi G_{\rm N} (T_{\mu\nu} +t_{\mu\nu} ),
 \label{eq4107}
\end{equation}
where the  term $t_{\mu\nu}$ on the right-hand side gives no contribution if 
the boundary spacetime represented by the metric $g^{(0)}_{\mu\nu}$ is conformally flat.

Now we  consider the consequences of this scenario for cosmology. 
To this end, we specify the boundary metric to have a  Friedmann-Robertson-Walker (FRW) form
\begin{equation}
ds^2=g^{(0)}_{\mu\nu}dx^\mu dx^\nu =dt^2 -a^2(t) d\Omega_k^2 .
 \label{eq3201}
\end{equation}
where
\begin{equation}
d\Omega^2_k=d\chi^2+\frac{\sin^2(\sqrt{k}\chi)}{k}(d\vartheta^2+\sin^2 \vartheta d\varphi^2)
\label{eq1004}
\end{equation}
is the spatial line element for a
closed ($k=1$), open hyperbolic ($k=-1$), or open flat ($k=0$) space. 
Next, we assume
\begin{equation}
{T}^{\mu}_{\nu}=\mbox{diag}(\rho,- p,-p,-p)
\label{eq3010}
\end{equation}
and using the effective Einstein equations (\ref{eq4107})  we obtain the holographic
Friedmann equations \cite{apostolopoulos,bilic1}
\begin{equation}
 H^2 +\frac{k}{a^2}-\frac{\ell^2}{4}\left(H^2 +\frac{k}{a^2}\right)^2=
 \frac{8\pi G_{\rm N}}{3}\rho,
 \label{eq3110}
\end{equation}
\begin{equation}
\left( \dot{H} -\frac{k}{a^2}\right) \left[1-\frac{\ell^2}{2}\left(H^2 +\frac{k}{a^2}\right)\right]=
 -4\pi G_{\rm N}(p+\rho).
 \label{eq4111}
\end{equation}
 Equations (\ref{eq3110}) and (\ref{eq4111}) imply
\begin{equation}
\dot{\rho}+3H(p+\rho)=0,
 \label{eq3109}
\end{equation}
which also follows from energy-momentum conservation
${T^{\mu\nu}}_{;\nu}=0$.

\section{Perturbations in the holographic cosmology}
\label{perturbations}

Here we derive the spectra of the cosmological perturbations for the
holographic cosmology with matter in the form of general $k$-essence.
We shall closely follow J.\ Garriga and V.\ F.\ Mukhanov  \cite{garriga}
and adjust their formalism to account for the Holographic cosmological perturbations.

\subsection{Background equations}

In the following we consider a spatially flat background with Friedmann equations (\ref{eq3110}) and  (\ref{eq4111}).
The beginning of inflation is characterized by the so called slow-roll regime
with slow-roll parameters satisfying $\varepsilon_i\ll 1$.
We use the following recursive definition of the slow roll parameters
\cite{schwarz,steer}
\begin{equation}
\varepsilon_{i+1}=\frac{\dot{\varepsilon}_i}{H\varepsilon_i} ,
 \label{eq3214}
\end{equation}
starting with
\begin{equation}
\varepsilon_1= -\frac{\dot{H}}{H^2}.
 \label{eq3114}
\end{equation}

Next we assume that apart from the conformal field there is  matter on the holographic brane
described by
a general $k$-essence action
\begin{eqnarray}
S_{\rm matt} = \int d^{4}x\sqrt{-g}\mathcal{L}(X,\theta)
\end{eqnarray}
where
$\mathcal{L}=\mathcal{L} (X,\theta)$ is an arbitrary function of the field $\theta$ and the kinetic term $X$,
\begin{equation}
X\equiv g^{\mu\nu}\theta_{,\mu}\theta_{,\nu} .
\label{eq0201}
\end{equation}
The energy-momentum tensor associated with $S_{\rm matt}$ is given by
\begin{eqnarray}
T_{\mu\nu} \equiv \frac{2}{\sqrt{-g}}\frac{\delta S_{\rm matt}}{\delta g^{\mu\nu}}
= 2 {\cal L}_{,X} \partial_\mu \theta \partial_\nu \theta  - g_{\mu\nu} {\cal L}
\label{eq0072}
\end{eqnarray}
where the subscript ${,X}$ denotes a partial derivative with respect to $X$.

Formally, one may proceed by solving the $\theta$ field equation
in conjunction with Einstein's
equations. However, it proves 
advantageous to pursue the hydrodynamic picture.
Using standard notation based on the similarity with perfect fluids \cite{garriga} 
(see also \cite{Unnikrishnan:2010ag, Arroja:2010wy, Piattella:2013wpa} for 
further details on the $k$-essence fluid correspondence), the pressure $p$ and energy  energy density  $\rho$ are given by
  \begin{equation}
p=\mathcal{L}, \quad \rho= 2X{\mathcal{L}}_{,X}-{\mathcal{L}}.
\label{eq0008}
\end{equation}
Then, the energy-momentum tensor can be expressed  as
\begin{eqnarray}
 {T}_{\mu\nu} 
=( {p}+ {\rho}) u_\mu u_\nu - {p}g_{\mu\nu} ,
\label{eq0005}
\end{eqnarray}
where 
\begin{eqnarray}
u_{\mu} \equiv \frac{\partial_{\mu}\theta}{\sqrt{X}}.
\end{eqnarray}
The definition above implies that $u^{\mu}u_{\mu}=1$.

\subsection{Scalar perturbations and power spectrum}
\label{scalar}

Assuming a spatially flat background with line element (\ref{eq3201}) with $k=0$, we introduce
the perturbed line element in the Newtonian gauge
\begin{equation}
 ds^2=(1+2\Psi) dt^2-(1-2\Phi)a^2(t)(dr^2+r^2 d\Omega^2) . \label{eq0013}
\end{equation}
This perturbed spacetime is no longer conformally flat
so the perturbed Einstein equations will have a nonvanishing contribution of
the tensor $\delta t_{\mu\nu}$.
However, 
for the moment we  assume that the  contribution of $\delta t_{\mu\nu}$ is of higher order in  the slow roll parameter expansion
and we justify this assumption a posteriori. 
Inserting the above metric components in the field equations (\ref{eq4107}) and ignoring $\delta t_\mu^\nu$ we obtain

\begin{eqnarray}
\frac{2}{a^{2}}\Bigg( 1-\frac{\ell^2}{2}H^{2} \Bigg)\big(\nabla^{2}\Phi
-3a^{2}H(\dot{\Phi}+H\Psi) \big) &= 8\pi G_{\rm N}\delta T^{0}_{0}
\label{eq:holo00}
\\
\nonumber
\\
2\Bigg( 1-\frac{\ell^2}{2}H^{2} \Bigg)\partial_{i}(\dot{\Phi}+H\Psi) &= 8\pi G_{\rm N}\delta T^{0}_{i}
\label{eq:holo0i}
\end{eqnarray}
\begin{eqnarray}
-2\Big[ \ddot{\Phi}+ H(\dot{\Psi}+3\dot{\Phi}) + \Psi (3H^{2}+2\dot{H})+\frac{1}{2a^{2}}\nabla^{2}(\Psi-\Phi)   \Big]
\delta^{i}_{j} +\ell^{2}\Big[ H^{2}\ddot{\Phi} + (3H^{3}+2H\dot{H})\dot{\Phi}
\nonumber
\\
+H^{3}\dot{\Psi} + (3H^{4}+4H^{2}\dot{H})\Psi - \frac{1}{6a^{2}}(3H^{2}+5\dot{H})
\nabla^{2}\Phi + \frac{1}{6a^{2}}(3H^{2}+\dot{H})\nabla^{2}\Psi  \Big]\delta^{i}_{j}
\nonumber\\
-\frac{1}{a^{2}}\partial^{i}\partial_{j}(\Phi-\Psi) +
\frac{\ell^{2}}{2a^{2}}(H^{2}+\dot{H})\partial^{i}\partial_{j} (\Phi - \Psi) = 8\pi G_{\rm N}\delta T^{i}_{j} 
\label{eq:holoij}
\end{eqnarray}
The perturbed energy-momentum tensor components are found from Eq.~\eqref{eq0005} and coincide with those of a perfect fluid. 
Besides, following standard Newtonian gauge conventions, the coordinates are chosen such that $u^i$ is a first order perturbative quantity ($u^i = \delta u^i$). 
Hence, up to the first perturbative order,
\begin{equation}
\delta T^{0}_{0} = \delta\rho
\end{equation}
\begin{equation}
\delta T^{0}_{i} = (\rho+p)\delta u_{i},
\label{eq:T0i}
\end{equation}
\begin{equation}
\delta T^{i}_{j} = -\delta^{i}_{j}\delta p  .
\label{eq:Tij}
\end{equation}
Then, from the off-diagonal part of \eqref{eq:holoij}
\begin{equation}
\left( 1- \frac{\ell^{2}}{2}(H^{2}+\dot{H}) \right)\partial^{i}\partial_{j}(\Phi-\Psi)=0
\end{equation}
we read off the slip parameter 
\begin{equation}
\eta\equiv \frac{\Phi}{\Psi} = 1,
\label{eq:slip}
\end{equation}
as in general relativity (GR). 
Hence, in the following we can work in longitudinal gauge
with $\Psi=\Phi$.
The slip parameter  is defined in Fourier $\boldsymbol{k}$-momentum space but,
following the common practice \cite{amendola2},  we omit the dependence on $\boldsymbol{k}$ in  Eq.\ (\ref{eq:slip}).

Now, the procedure described in  Ref.~\cite{garriga} (see also the appendix of \cite{bilic3} for more details)
can be applied, keeping in mind that the background evolution is governed by
Eqs.\ (\ref{eq3110}) and  (\ref{eq4111}) with $k=0$.
The relevant Einstein equations at linear order are given by
\begin{equation}
a^{-2} \nabla^2 \Phi -3H\dot{\Phi} +3H^2\Phi=4\pi G_{\rm N}
\delta {T}^0_0 (1-h^2/2)^{-1},
 \label{eq0014}
\end{equation}
\begin{equation}
(\dot{\Phi}+H\Phi)_{,i}=4\pi G_{\rm N}
\delta {T}^0_i (1-h^2/2)^{-1},
 \label{eq0015}
\end{equation}
where we have abbreviated
\begin{equation}
h\equiv \ell H.
 \label{eq2015}
\end{equation}
 Equations (\ref{eq0014}) and (\ref{eq0015}) 
are sufficient for deriving the scalar power spectra.
However, to check the consistency of our assumptions,
we will also consider the $ij$ components of the Einstein
equations at linear order,
\begin{equation}\tag{35a}
h^2\left(\Phi\dot{H}+\frac{\dot{H}}{H}\dot{\Phi}-\frac{1}{3a^{2}}\frac{\dot{H}}{H^{2}}\Delta\Phi\right)\delta^{i}_{j}
	-\left(\ddot{\Phi}+4H\dot{\Phi}+2\Phi\dot{H}+3H^{2}\Phi\right)
	\left(1-\frac{h^2}{2}\right)\delta^{i}_{j}
	=4\pi G_{\rm N}\delta T^{i}_{j}.
	\label{eqTij}
\end{equation}
The  perturbations of  the stress tensor components $\delta {T}^\mu_\nu$ are induced by the perturbations of
the scalar field $\theta(t,x)=\theta(t)+\delta\theta(t,x)$ and by the metric perturbation \cite{garriga} 
(see also \cite{mukhanov3,amendola1}). 
Keeping up with the fluid analogy we define the speed of sound
\begin{equation}
c_{\rm s}^2\equiv \left.\frac{\partial p}{\partial\rho}\right|_\theta =\frac{p_{,X}}{\rho_{,X}}
=\frac{p_{,X}}{p_{,X}+2Xp_{,XX}}
=\frac{p+\rho}{2X \rho_{,X}}  .
\label{eq0018}
\end{equation}
With this, the component $\delta T_0^0$ at first perturbative order can be written as
\begin{equation}
	\delta T^0_0 =  \rho_{,X} \delta X+\rho_{,\theta }	\delta \theta 
	= \frac{\rho + p}{2 X c^2_s} \left ( \delta g^{\mu \nu} \partial_\mu \theta \partial_\nu \theta 
+ 2 g^{\mu \nu} \partial_{\mu} \theta \partial_\nu \delta \theta -\frac{\dot X}{\dot \theta } \delta \theta \right) 
+ \frac{\dot \rho}{\dot \theta }\delta \theta \, .
\label{eq0071}
\end{equation} 

Since $k$-essence is  described by an action, diffeomorphism invariance implies that 
the energy-momentum tensor associated with the action must be conserved. 
At background level the conservation equation in fluid notation takes the usual form given by Eq.~\eqref{eq3109}.
Using this and the fact that $X = \dot \theta^2$ at background level,
from (\ref{eq0071})
we obtain 
\begin{equation}
\delta {T}^0_0=\frac{{p}+{\rho}}{{c}_{\rm s}^2}
\left[\left(\frac{\delta\theta}{\dot{\theta}}\right)^.-\Phi\right]
-3H({p}+{\rho}) \frac{\delta\theta}{\dot{\theta}}.
 \label{eq0041}
\end{equation}
Similarly,  from \eqref{eq:T0i} we obtain
\begin{equation}
	\delta {T}^0_i=({p}+{\rho})\left(\frac{\delta\theta}{\dot{\theta}}\right)_{,i} \, ,
\label{eq0043}
\end{equation}
Equations (\ref{eq0018}-\ref{eq0043}) are no novelty \cite{garriga}, and are reviewed here only for convenience. 
For the $ij$ components, Eq.\  \eqref{eq:Tij} combined with (\ref{eq0041}) yields 
\begin{equation}\tag{39a}
	\delta T^{i}_{j}=- c_{\rm s}^{2}\delta T^{0}_{0}\delta^{i}_{j}+(c_{\rm s}^{2} \rho_{,\theta}-p_{,\theta})\delta\theta \delta^{i}_{j}.
	\label{Tij}
\end{equation}
In particular, for a pure $k$-essence the second term on the right hand side of (\ref{Tij}) vanishes in which case we have 
\begin{equation}\tag{39b}
	\delta T^{i}_{j}=- c_{\rm s}^{2}\delta T^{0}_{0}\delta^{i}_{j} .
	\label{Tij2}
\end{equation}

In the above derivation it was not necessary to explicitly
employ  the field equation obtained from the variation with respect to $\theta$ because,
as we have demonstrated in Appendix \ref{appendix}, the equation of motion of the $\theta$ field is equivalent 
to the energy momentum conservation law. 
Therefore, the fluid-based notation, as used above, 
is fully equivalent to using the field equations derived from the fundamental fields $\theta$ and $g_{\mu \nu}$. 
Alternatively, 
we could have derived Eq.\ (\ref{eq0041}) and (\ref{eq0043})
starting directly from the definition of the energy-momentum tensor (\ref{eq0072})
and employing the equation of motion of the $\theta$ field, without resorting to the hydrodynamic  description.
We show this in Appendix \ref{appendixB}.

In the slow roll regime the sound speed deviates slightly from unity and may be expressed in terms of
slow-roll parameters.
First, by making use of the definition (\ref{eq3114}) and modified Friedmann equations (\ref{eq3110}) and  (\ref{eq4111})
with (\ref{eq0008}),
we can express
the variable $X$ in the slow roll regime as
\begin{equation}
X=- \frac{2p(2-h^2)}{3p_{,X}(4-h^2)}\varepsilon_1+ \mathcal{O}(\varepsilon_i^2),
\label{eq5107}
\end{equation}
Then from (\ref{eq0018}) we find
\begin{equation}
c_{\rm s}^2=1+ \frac{4(2-h^2)}{3(4-h^2)}\frac{p p_{,XX}}{p_{,X}^2}\varepsilon_1+ \mathcal{O}(\varepsilon_i^2).
\label{eq5108}
\end{equation}
For example, in the tachyon model with
Lagrangian $\mathcal{L}=-V\sqrt{1-X}$ one finds \cite{bilic3}
\begin{equation}
c_{\rm s}^2=1-\frac{4(2-h^2)}{3(4-h^2)}\varepsilon_1+ \mathcal{O}(\varepsilon_i^2).
\label{eq5109}
\end{equation}

Using (\ref{eq0041}) and (\ref{eq0043}) equations (\ref{eq0014}) and (\ref{eq0015})
take the form
\begin{equation}
  \left(\frac{\delta\theta}{\dot{\theta}}\right)^{\mbox{.}}
=\Phi + \frac{c_{\rm s}^2}{4\pi G_{\rm N}a^2(\tilde{p}+\tilde{\rho})}\nabla^2 \Phi ,
 \label{eq0033}
\end{equation}
\begin{equation}
(a\Phi)^{\mbox{.}}=4\pi G_{\rm N} a(\tilde{p}+\tilde{\rho})\frac{\delta\theta}{\dot{\theta}},
 \label{eq0034}
\end{equation}
where we have defined 
\begin{equation}
\tilde{p}+\tilde{\rho}=(p+\rho)(1-h^2/2)^{-1}.
 \label{eq1034}
\end{equation}

To check the consistency of our assumptions, we have examined the $ii$ components 
 of  Einstein's equations. Using Eqs.\ (\ref{eqTij}) and (\ref{Tij2}) for $i=j$ we have derived an equation similar to (\ref{eq0033}).
 Combining with (\ref{eq0014}), (\ref{eq0015}), and (\ref{eq0041}), we have obtained
\begin{equation}\tag{45a}
	\left(\frac{\delta\theta}{\dot{\theta}}\right)^{\mbox{.}}
	=\Phi + \frac{c_{\rm s}^2}{4\pi G_{\rm N}a^2(\tilde{p}+\tilde{\rho})}
\left(1+\frac{h^2}{3(1-h^2/2)c_{\rm s}^2}\frac{\dot{H}}{H^2}\right)
	\nabla^2 \Phi ,
	\label{eq1033}
\end{equation} 
which departs from  (\ref{eq0033}) by an addition of the order of $\varepsilon_1$. Clearly, this discrepancy is a consequence of neglecting the
contribution of $\delta t_{\mu\nu}$ terms in the perturbed Einstein equations
(\ref{eq:holo00})-(\ref{eq:holoij}). As we have anticipated, this analysis justifies our assumption that the  contribution of $\delta t_{\mu\nu}$ will be of higher order in  the slow roll parameter expansion. In the rest of the paper
we will use Eq.\ (\ref{eq0033}) to calculate the power spectra at the lowest orders 
in epsilon parameters.

So far we have  not used the
modified Friedmann  cosmology so
equations (\ref{eq0033}) and (\ref{eq0034})
will coincide with those derived in \cite{garriga}
if $p+\rho$ in \cite{garriga} is replaced by $\tilde{p}+\tilde{\rho}$.
We now use the equation
\begin{equation}
\dot{H}= -4\pi G_{\rm N} (\tilde{p}+\tilde{\rho}).
 \label{eq2034}
\end{equation}
which follows from (\ref{eq4111}).
As in Ref.~\cite{garriga}, we introduce  new functions
\begin{equation}
\xi=\frac{a\Phi}{4\pi G_{\rm N}H}, \quad
 \zeta = \Phi+H\frac{\delta\theta}{\dot{\theta}}.
\label{eq0021}
\end{equation}
 The quantity $\zeta$ is gauge invariant and  measures the spatial curvature of comoving
(or constant-$\theta$) hyper-surfaces.
Substituting the definitions (\ref{eq0021}) into (\ref{eq0033}) and (\ref{eq0034}) and using
(\ref{eq2034})
we find 
\begin{equation}
a\dot{\xi}=z^2c_{\rm s}^2\zeta,
 \label{eq0016}
\end{equation}
\begin{equation}
a\dot{\zeta}=z^{-2}\nabla^2\xi,
 \label{eq0017}
\end{equation}
where
\begin{equation}
z=\frac{a(\tilde{p}+\tilde{\rho})^{1/2}}{c_{\rm s}H} =
\frac{a}{c_{\rm s}}\sqrt{\frac{\varepsilon_1}{4\pi G_{\rm N}}} .
 \label{eq0029}
\end{equation}
Here we have used the definition (\ref{eq3114}) of the slow roll parameter $\varepsilon_1$.
Equations (\ref{eq0016}) and (\ref{eq0017})  with (\ref{eq0029}) are identical to those obtained in the standard 
$k$-inflation
\cite{garriga}.

 Note that the only change  in Eqs.~(\ref{eq0016}) and (\ref{eq0017}) compared with those in Ref. \cite{garriga} 
 comes from the factor $1 - h^2/2$ that multiplies $\rho + p$ inside $z$ per Eq.~\eqref{eq0029}. 
However, this factor is absorbed into $\varepsilon_1$ due to Eq.~\eqref{eq2034}. 
As a result, the second equality in Eq.\ \eqref{eq0029} is exactly the same as in Ref. \cite{garriga}.   
This feature is due to the fact that the factor  $1 - h^2/2$ appears  on the right-hand side of 
the perturbation equations (\ref{eq0016}) and (\ref{eq0017}) in the same way as in the background 
evolution equation (\ref{eq2034}).
In contrast, in Ref.\ \cite{bilic3}, where 
an approximate scheme was used, 
the expression for $z$  as a function of $\varepsilon_1$, explicitly 
contains the factor $1-h^2/2$, thus leading to different equations than those of Ref.~\cite{garriga}.

By introducing the conformal time $\tau=\int dt/a$ and a new variable $v=z\zeta$, it is straightforward
to show from equations
(\ref{eq0016}) and (\ref{eq0017}) that  $v$ satisfies a second order differential equation
\begin{equation}
v''-c_{\rm s}^2 \nabla^2 v-\frac{z''}{z}v =0,
 \label{eq0022}
\end{equation}
where the primes denote derivatives with respect to $\tau$.
By making use of the Fourier transformation,
we also obtain the  mode-function equation
\begin{equation}
v_q''+\left(c_{\rm s}^2q^2  -\frac{z''}{z}\right)v_q =0.
 \label{eq0032}
\end{equation}
As we are looking for a solution to this equation in the slow-roll regime, it is useful to express
the quantity $z''/z$ in terms of slow-roll parameters $\varepsilon_i$.
In the slow-roll regime one  can use  the relation
\cite{lidsey1}
\begin{equation}
\tau=- \frac{1+\varepsilon_1}{aH}+\mathcal{O}(\varepsilon_1^2),
\label{eq0047}
\end{equation}
which  follows from the definition (\ref{eq3114}) expressed in terms of the conformal time.
At linear order in $\varepsilon_i$ we find
\begin{equation}
\frac{z''}{z}=\frac{\nu^2-1/4}{\tau^2},
\label{eq0050}
\end{equation}
where
\begin{equation}
\nu^2=\frac94+
3\varepsilon_1+\frac32\varepsilon_2  .
\label{eq0048}
\end{equation}

We look for
a solution to (\ref{eq0032}) which satisfies the positive frequency asymptotic limit
\begin{equation}
\lim_{\tau\rightarrow -\infty}v_q=\frac{e^{-ic_{\rm s}q\tau}}{\sqrt{2c_{\rm s}q}}.
 \label{eq0027}
\end{equation}
Then the solution which up to a phase agrees with (\ref{eq0027})
is
\begin{equation}
v_q=\frac{\sqrt{\pi}}{2}(-\tau)^{1/2} H_\nu^{(1)}(-c_{\rm s}q\tau),
 \label{eq0049}
\end{equation}
where $H_\nu^{(1)}$ is the Hankel function of the first kind of rank $\nu$.

 It is of interest to show explicitly that the perturbation of the scalar field $\theta$ and metric perturbation $\Phi$ are smooth functions of time.
The perturbations $\delta{\theta}$ and $\Phi$ are  encoded in the functions 
$\zeta$ and $\xi$ that satisfy the first order differential equations (\ref{eq0016}) and (\ref{eq0017}).
These equations are equivalent to a single second order differential equation (\ref{eq0022}) for the function
$v$. Hence, the functions $\zeta$ and $\xi$ can be expressed in terms of $v$ and its derivative $\dot{v}$.
This in turn implies that the perturbations $\delta{\theta}$ and $\Phi$ can be expressed in terms of $v$ and $\dot{v}$.
More explicitly,  from Eqs.\ (\ref{eq0021})-(\ref{eq0017}) it follows in momentum space
\begin{eqnarray}
\delta\theta= \left[
\left(\frac{1}{zH}-\frac{4\pi G_{\rm N}\dot{z}}{q^2}   \right)v_q
+\frac{4\pi G_{\rm N}z}{q^2}\dot{v}_q 
\right]\dot{\theta}, 
\end{eqnarray}
\begin{eqnarray}
\Phi= 
\frac{4\pi G_{\rm N}}{q^2}(\dot{z}v_q-z\dot{v}_q),
\end{eqnarray}
where $v_q$ is a solution to (\ref{eq0032}) in momentum space. In the slow-roll regime the solution is
expressed in terms of the Hankel functions of the first kind (\ref{eq0049}).
This type of solution appears in all scalar models of inflation (see, e.g., \cite{baumann}). From our solutions depicted in Fig.~\ref{fig1}, the slope $\dot{\theta}$  is everywhere finite 
and asymptotically approaches a finite constant or zero.
Hence,  there are reasonable $k$-essence models that yield both $\Phi$ and $\theta$ as smooth functions in the domain of inflation.

 Applying the standard canonical quantization \cite{mukhanov}  the field $v_q$ is promoted to an operator
 and the power spectrum of the field $\zeta_q=v_q/z$ is obtained  from the two-point correlation function
\begin{equation}
\langle\hat{\zeta}_q\hat{\zeta}_{q'}\rangle=\langle\hat{v}_q \hat{v}_{q'}\rangle/z^2=
(2\pi)^3 \delta(\mbox{\boldmath $q$}+\mbox{\boldmath $q$}')|\zeta_q|^2.
 \label{eq0063}
\end{equation}
The 
spectral density
\begin{equation}
\mathcal{P}_{\rm S}(q)=\frac{q^3}{2\pi^2}|\zeta_q|^2=\frac{q^3}{2\pi^2z^2}|v_q|^2 ,
 \label{eq0024}
\end{equation}
with $v_q$ given by (\ref{eq0049}),
characterizes the primordial scalar fluctuations.

Next,
  we evaluate the scalar spectral density at the horizon crossing, i.e., for  a wave-number satisfying $q=aH$.
Following Refs.\ \cite{steer,hwang} we make use of the expansion of the Hankel function
for $c_{\rm s}q\tau \ll 1$,
\begin{equation}
 H_\nu^{(1)}(-c_{\rm s}q\tau)\simeq -\frac{i}{\pi}\Gamma(\nu)\left(\frac{-c_{\rm s}q\tau}{2}\right)^{-\nu},
 \label{eq0053}
\end{equation}
where the conformal time $\tau<0$ and $q$ is the comoving wave number.
Using this
we find  at the lowest order in $\varepsilon_1$ and $\varepsilon_2$
\begin{equation}
 \mathcal{P}_{{\rm S}} \simeq \frac{G_{\rm N} H^2 }{\pi c_{\rm s}\varepsilon_1 }
 \left[1-2\left(1+C\right)\varepsilon_1-C \varepsilon_2\right],
\label{eq3007}
\end{equation}
 where $C=\gamma-2+\ln 2 \simeq -0.73$ and $\gamma$ is the Euler constant,
so we recover the standard expression \cite{steer,hwang}. 
However, it should be stressed that the relation between the sound speed and $\varepsilon_i$ is not the usual one, 
as shown in Eq.~\eqref{eq5108}. 
Hence, in our case the contribution of the sound speed
will be altered in comparison with the standard result.

\subsection{Tensor perturbations}
\label{tensor}

The tensor perturbations are related to the production of gravitational waves
during inflation.
The metric perturbation is, in this case, written as
\begin{equation}
ds^2= dt^2-a^2(t)\left(\delta_{ij}+h_{ij}\right)dx^idx^j,
\label{eq0054}
\end{equation}
where $h_{ij}$ is traceless and transverse.
Inserting the metric components in the field equations (\ref{eq4107}) we obtain
\begin{equation}
 \left(1-\frac{\ell^2}{2}(H^2+\dot{H})\right)
 \left(\ddot{h}_{ij}+3H \dot{h}_{ij}-\frac{1}{a^2}\nabla^2 h_{ij}\right)=8\pi G_{\rm N} \delta T_{ij}.
 \label{eq2055}
\end{equation}
In the absence of anisotropic stress, the gravitational waves are decoupled from matter and  the right-hand side of 
the above equation is zero. 
Hence, our braneworld scenario 
does not introduce changes to the gravitational waves dynamics.
The 
spectral density which characterizes the primordial tensor fluctuations, 
in the approximation $q\tau \ll 1$, is given by
\begin{equation}
 \mathcal{P}_{{\rm T}} \simeq \frac{16 G_{\rm N} H^2 }{\pi}
 \left[1-2\left(1+C\right)\varepsilon_1\right].
\label{eq4007}
\end{equation}
Hence, the tensor perturbation spectrum 
 is given by the usual expression  \cite{lidsey1}
in the same way as our scalar spectrum $\mathcal{P}_{\rm S}$.

\subsection{Scalar spectral index and tensor to scalar ratio}
\label{index}

The scalar spectral index $n_{\rm S}$ and tensor to scalar ratio $r$ are given by
\begin{equation}
n_{\rm S}-1= \frac{d\ln \mathcal{P}_{{\rm S}}}{d\ln q}\simeq\frac{1}{H(1-\varepsilon_1)}\frac{d\ln \mathcal{P}_{{\rm S}}}{dt},
\label{eq3006}
\end{equation}
\begin{equation}
r=\frac{\mathcal{P}_{\rm T}}{\mathcal{P}_{{\rm S}}},
\label{eq3005}
\end{equation}
where $\mathcal{P}_{{\rm S}}$ and $\mathcal{P}_{\rm T}$ are
evaluated at the horizon crossing with $q=a H$.
From (\ref{eq3007}) and (\ref{eq4007}) keeping the terms up to the quadratic order in $\varepsilon_i$
we obtain
\begin{equation}
r=16 \varepsilon_1 \left[1+C\varepsilon_2
+\frac{2(2-h^2)}{3(4-h^2)}\frac{p p_{,XX}}{p_{,X}^2}\varepsilon_1\right]
\label{eq128}
\end{equation}
and
\begin{eqnarray}
&& n_{\rm s}=1-2\varepsilon_1-\varepsilon_2
-\left(2+\frac{8h^2}{3(4-h^2)^2}\frac{p p_{,XX}}{p_{,X}^2}\right)\varepsilon_1^2
\nonumber\\
&& -\left(3+2C +\frac{2(2-h^2)}{3(4-h^2)}\frac{p p_{,XX}}{p_{,X}^2}\right)\varepsilon_1\varepsilon_2
- C \varepsilon_2\varepsilon_3.
 \label{eq129}
\end{eqnarray}

Note that our results 
at linear order in $\varepsilon_i$ agree with the standard scalar field inflation as well as with the general 
$k$-essence inflation \cite{hwang}.
At quadratic order
the standard $k$-inflation results
are recovered in the limit $\ell\rightarrow 0$ ($h\rightarrow 0$).  
For example,
if we specify the $k$-essence Lagrangian to the tachyon,
in the limit $\ell\rightarrow 0$ we recover the standard tachyon inflation (e.g., compare with Ref.\ \cite{steer}).

\section{Summary and conclusions}
We have studied the early cosmology on the holographic brane
where the effective Einstein equations are modified due to the dual conformal theory
on the AdS boundary.
We have developed  the complete perturbation theory at linear order for 
gravity on the holographic brane  together with a general $k$-essence field.

We  have derived
 the  scalar and tensor power spectra  up to the second order in the slow-roll parameter
 expansion
 and calculated the scalar spectral index $n_{\rm s}$  and scalar to tensor ratio $r$.
We have found that 
the expressions for both the scalar and tensor power spectra have the same form as those in
the standard $k$-inflation. 
This is a remarkable result in view of quite nontrivial modification of 
Einstein's equations in the holographic braneworld.
However, the functional dependence of the sound speed on 
the slow-roll parameters is altered and, as a result, the second order terms
in the slow-roll parameter expansion of $n_{\rm s}$  and $r$ 
are modified compared to those in the standard $k$-inflation, as shown in Eqs.\ (\ref{eq128}) and (\ref{eq129}).
 
It would be of considerable interest to investigate 
the holographic inflation for a particular $k$-essence model or any other interesting model
with a nontrivial sound speed. However, this would go beyond the scope of the present paper
as our purpose here is to show how the holographic braneworld scenario affects
inflation driven by a general $k$-essence.

\section*{Acknowledgments}
N.R.~Bertini thanks CAPES (Brazil) for support. The work of N.~Bili\'c  has been partially supported by
the European Union through the European Regional Development Fund - the Competitiveness and
Cohesion Operational Programme (KK.01.1.1.06).
D.C.~Rodrigues thanks  CNPq (Brazil) and FAPES (Brazil) for partial support.
This study was financed in part by the 
{\it Coordena\c{c}\~ao de Aperfei\c{c}oamento de Pessoal de N\'ivel Superior - Brasil} (CAPES) - Finance Code 001.

\appendix

\section{Energy conservation and equation of motion} 
\label{appendix}

Consider the $k$-essence action
\begin{eqnarray}
S_{m} = \int d^{4}x\sqrt{-g}\mathcal{L}(X,\theta)
\label{eq1}
\end{eqnarray}
where
\begin{eqnarray}
X = g^{\mu\nu}\partial_{\mu}\theta \partial_{\nu}\theta.
\label{eq2}
\end{eqnarray}
In the cosmological context it is appropriate to assume $X>0$.
Variation of (\ref{eq1}) yields  the equation of motion
\begin{eqnarray}
\left(2\mathcal{L}_{,X} g^{\mu\nu}\theta_{,\nu}\right)_{;\mu}- \mathcal{L}_{,\theta} =0
\label{eq3}
\end{eqnarray}
where we  denote
\begin{eqnarray}
\mathcal{L}_{,X}\equiv\frac{\partial \mathcal{L}}{\partial X}, \quad\quad
 \mathcal{L}_{,\theta}\equiv  \frac{\partial \mathcal{L}}{\partial \theta} .
\label{eq4}
\end{eqnarray}
The energy-momentum tensor associated with $S_{\rm matt}$ 
can be expressed in the perfect fluid form
\begin{equation}
 {T}_{\mu\nu}=( {p}+ {\rho}) u_\mu u_\nu - {p}g_{\mu\nu}.
\label{eq13}
\end{equation}
where we use the hydrodynamic variables  
\begin{eqnarray}
p = \mathcal{L},
\quad\quad
\rho =  2X  \mathcal{L}_{,X} - \mathcal{L}_{,\theta},
\label{eq6}
\end{eqnarray}
and 
\begin{eqnarray}
u_{\mu} = \frac{\partial_{\mu}\theta}{\sqrt{X}}
\label{eq5}
\end{eqnarray}

The  energy-momentum conservation equation
\begin{equation}
 {{T}^{\mu\nu}}_{;\nu}=0
\end{equation}
contracted with $u^\mu$ and applied to (\ref{eq13}) yields 
another expression of the energy  conservation 
\begin{equation}
\dot{\rho}+3H (p+\rho)=0,
\label{eq0}
\end{equation}
where $\dot{\rho}\equiv u^\mu \rho_{,\mu}$ and the expansion rate 
$H={u^\mu}_{;\mu}/3$.

For a general $k$-essence theory  one can proof the following theorem:
{\em Equation of motion for a $k$-essence field is satisfied if and only if ${T^{\mu\nu}}_{;\nu}=0$}.
For our purpose it is sufficient to prove that the 
equation of motion (\ref{eq3}) is equivalent to
the energy conservation equation (\ref{eq0})
provided $X>0$.

{\bf Proof} \\
Using (\ref{eq5}) Eq.\ (\ref{eq3}) can be written as
\begin{eqnarray}
(2\sqrt{X} \mathcal{L}_{,X} u^\mu)_{;\mu}- \mathcal{L}_{,\theta} =0.
\label{eq7}
\end{eqnarray}
Multiplying  by $\sqrt{X}$, equation (\ref{eq7})  can be recast in the form
\begin{eqnarray}
2X\mathcal{L}_{,X} {u^\mu}_{;\mu} +\mathcal{L}_{,X}u^\mu X_{,\mu}+2X u^{\mu}(\mathcal{L}_{,X})_{,\mu}
- \sqrt{X}\mathcal{L}_{,\theta} =0.
\label{eq8}
\end{eqnarray}
Then by making use of  the Leibniz rule and (\ref{eq5}) we obtain
\begin{eqnarray}
2X\mathcal{L}_{,X} {u^\mu}_{;\mu} + u^{\mu}(2X\mathcal{L}_{,X})_{,\mu} -\mathcal{L}_{,X}u^\mu X_{,\mu}
- u^\mu\theta_{\mu}\mathcal{L}_{,\theta} =0.
\label{eq9}
\end{eqnarray}
Assembling the last two terms we obtain
\begin{eqnarray}
2X\mathcal{L}_{,X} {u^\mu}_{;\mu} + u^{\mu}(2X\mathcal{L}_{,X}-\mathcal{L})_{,\mu}  =0.
\label{eq10}
\end{eqnarray}
Using (\ref{eq6}) this can be written as  
\begin{equation}
{u^\mu}_{;\mu} (p+\rho)+u^\mu\rho_{,\mu}=0
\label{eq11}
\end{equation}
Identifying ${u^\mu}_{;\mu}=3H$ and $u^\mu\rho_{,\mu}=\dot{\rho}$
 we obtain the energy conservation  equation  in the usual form (\ref{eq0}).
 Hence, we have shown that the equation of motion (\ref{eq7}) implies the energy conservation equation (\ref{eq11}).
The reverse statement is also true since every step in our derivation from (\ref{eq7})  to (\ref{eq11}) 
is reversible and hence,  equations (\ref{eq7}) and (\ref{eq11}) are equivalent.

In a cosmological background described by the metric (\ref{eq3201}) we have $X=\dot{\theta}^2$.
Then, the equation of motion (\ref{eq3}) takes the form  
\begin{eqnarray}
2(2X\mathcal{L}_{,XX}+\mathcal{L}_{,X})\ddot{\theta}+6H\mathcal{L}_{,X}\dot{\theta}
+2X\mathcal{L}_{,X\theta} -\mathcal{L}_{,\theta}  =0.
\label{eq12}
\end{eqnarray}
It may be easily verified that the same equation is obtained also from the 
energy conservation equation (\ref{eq0}).

\section{Alternative derivation of the energy-momentum perturbations} 
\label{appendixB}

Here we derive the perturbations of the energy momentum tensor directly from it's definition 
\begin{equation}
T^{\mu}_{\nu} = 2 {\cal L},_{X} g^{\mu\alpha}\partial_{\alpha}\theta \partial_{\nu}\theta - \delta^{\mu}_{\nu}{\cal L}.
\end{equation}
At linear order we find
\begin{eqnarray}
\delta T^{\mu}_{\nu} = 2 ({\cal L},_{XX}\delta X + {\cal L},_{X\theta}\delta\theta)g^{\mu\alpha}\partial_{\alpha}\theta 
\partial_{\nu}\theta + 2{\cal L},_{X}(\delta g^{\mu\alpha}\partial_{\alpha}\theta\partial_{\nu}\theta 
+ 2g^{\mu\alpha}\partial_{\alpha}\theta\partial_{\nu}\delta\theta) \nonumber \\
- \delta^{\mu}_{\nu}({\cal L},_{X}\delta X  +  {\cal L},_{\theta}\delta \theta) 
\end{eqnarray}
and owing to
\begin{equation}
\delta X = \delta g^{\mu\nu}\partial_{\mu}\theta\partial_{\nu}\theta + 2g^{\mu\nu}\partial_{\mu}\theta\partial_{\nu}\delta\theta
= -2 X\Phi + 2\dot{\theta} \delta \dot{\theta},
\end{equation}
we obtain
\begin{equation}
\delta {T}^0_0=(2X\mathcal{L}_{,XX} +\mathcal{L}_{,X})(2 \dot{\theta}\delta\dot{\theta}-2X\Phi)
+(2X\mathcal{L}_{,X\theta} -\mathcal{L}_{,\theta})\delta{\theta}
 \label{eq0073}
\end{equation}
and 
\begin{equation}
\delta {T}^0_i=(2X\mathcal{L}_{,X})\left(\frac{\delta\theta}{\dot{\theta}}\right)_{,i}  .
 \label{eq0074}
\end{equation}
Then, combining Eq. (\ref{eq0073}) with $\theta$ field equation (\ref{eq12}) derived in Appendix \ref{appendix},
we obtain
\begin{equation}
\delta {T}^0_0=2(2X\mathcal{L}_{,XX} +\mathcal{L}_{,X})( \dot{\theta}\delta\dot{\theta}
-\ddot{\theta}\delta\theta-X\Phi)
-6H \mathcal{L}_{,X} \dot{\theta} \delta{\theta}.
 \label{eq0075}
\end{equation}
It may be easily verified that  Eqs.\ (\ref{eq0075}) and (\ref{eq0074}) will be identical to
Eqs.\ (\ref{eq0041}) and (\ref{eq0043}), respectively,  if the fluid variables $p$, $\rho$, and $c_{\rm s}^2$ 
in (\ref{eq0041}) and (\ref{eq0043})
are replaced by their original field theory expressions. 


\section{Background evolution of the field in particular $k$-essence models}
\label{appendixC}

In order to show the behavior of the $\theta$ field and for illustration purposes, we study here a few examples of $k$-essence
in the holographic braneworld scenario.
We consider two classes of $k$-essence: the canonical scalar field and tachyon condensate models. 
The Lagrangian density for the canonical case is given by
\begin{eqnarray}
{\cal L} = \frac 12  X  - V(\theta) ,
 \label{eq:canonical}
\end{eqnarray}
and for the tachyon condensate by \cite{sen}
\begin{eqnarray}
{\cal L} = - V(\theta)\sqrt{1-X}.
\label{eq:tac}
\end{eqnarray}
In order to proceed with numerical calculations, we need to specify $V(\theta)$. 
For the canonical model we make use of 
\begin{equation}
	V(\theta) = \frac{1}{2}m^{2}\theta^{2} \, . 
\label{eq:canonicalmass}
\end{equation}
as the simplest potential being used for the large field or chaotic inflation \cite{Martin:2013tda,baumann}
For the tachyon condensate we take the exponential potential which has been extensively exploited in the tachyon literature
\cite{steer,sami,cline,bilic3}
\begin{equation}
	V(\theta)  = \ell^{-4} e^{-\omega \theta} 
\label{eq:tacExp}
\end{equation}
and the inverse power-law potential \cite{brax,abramo,bilic4}
\begin{equation}
	V(\theta)  = m^{4-n} \theta^{-n}, \quad n>0.   
\label{eq:tacPower}	
\end{equation}

We start from the $\theta$-field equation \eqref{eq12} together with the expansion rate  $h \equiv \ell H $
expressed in terms of $\theta$. From Eq.\ (\ref{eq3110}) with $k=0$ we obtain 
\begin{eqnarray}
h^{2} = 2\left(1 - \sqrt{1- \frac{8 \pi G_{\rm N}}{3}\ell^{2}\rho} \right), 
 \label{eq:2}
\end{eqnarray}
as a solution to a quadratic equation that meets 
 the requirement that $H$ should be zero for $\rho=0$. 
Initial conditions are taken at $t=0$ where we fix $h_{\rm i}=\sqrt2$ and set
$\dot{\theta}(0)=0$ in all examples. The initial   $\theta_0\equiv \theta(0)$ is then fixed by the first Friedmann equation.

\begin{figure}[hbt]
\includegraphics[width=0.95\textwidth]{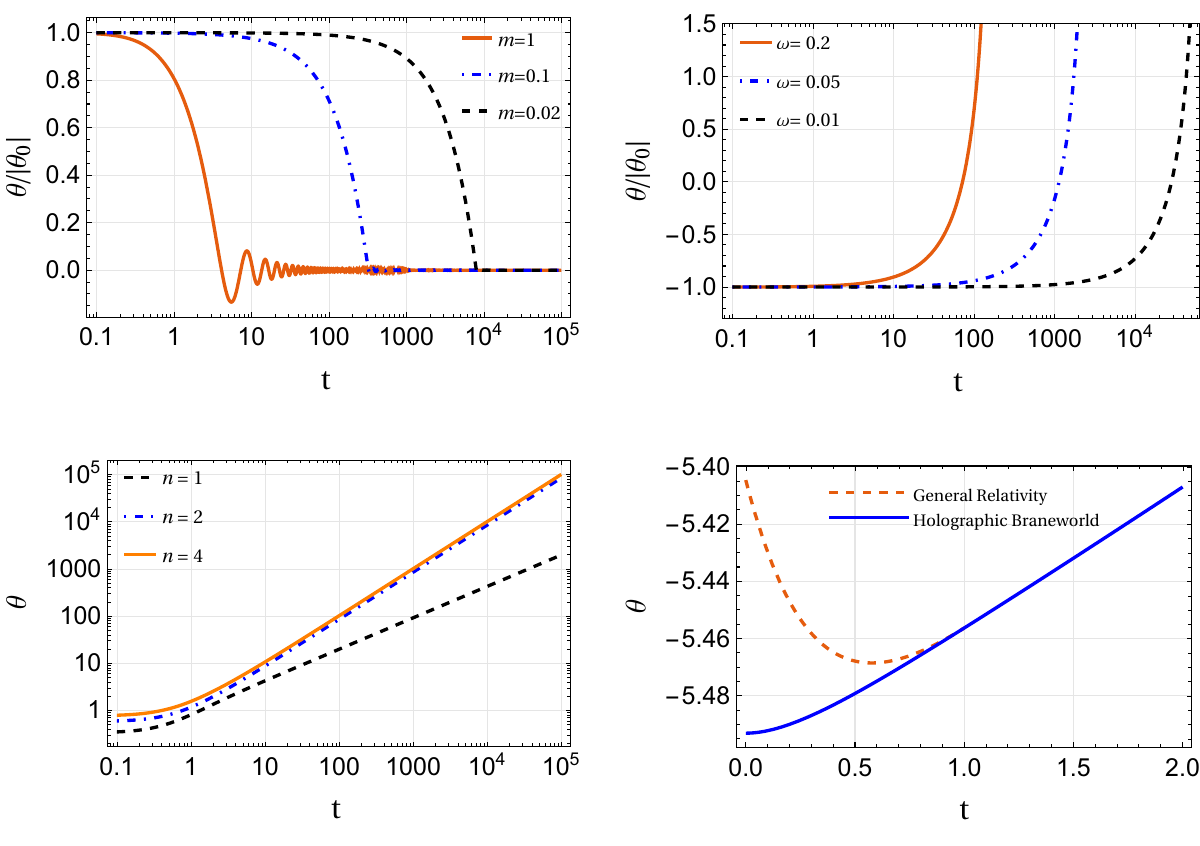}
\caption{Plots of the  field $\theta$ versus time (in units of $\ell$) for different  choices of $k$-essence.
Upper left: canonical scalar field with quadratic potential (\ref{eq:canonicalmass}); 
upper right: tachyon $k$-essence field with exponential potential (\ref{eq:tacExp});   
lower left: tachyon $k$-essence with inverse power-law potential (\ref{eq:tacPower});
lower right: comparison between the holographic braneworld and GR scenarios.
The field $\theta$ in the canonical model, the mass $m$, and the parameter $\omega$  are in units of $\ell^{-1}$,
whereas
$\theta$ in the tachyon model is in units of $\ell$.
}
  \label{fig1}
\end{figure}

In Fig.\ \ref{fig1} we depict  our numerical results as four plots  of $\theta$ as a function of time.
The canonical field, as expected, vanishes asymptotically with damped oscillations whereas the tachyon field grows
linearly for large $t$. 
In the lower right panel we compare the tachyon model with exponential potential in 
the holographic braneworld and in general relativity. The initial condition for the GR 
model is adjusted so that two models agree in the low density limit, i.e., at large $t$.
These two scenarios disagree substantially only at small $t$, as expected.

\end{document}